\documentclass[prd,twocolumn,nofootinbib]{revtex4}

\usepackage{mathrsfs}
\usepackage{subfigure}
\usepackage{amssymb}
\usepackage{amsmath}
\usepackage{epsfig}
\usepackage{float}
\usepackage{caption}
\usepackage{color}
\usepackage{upgreek}
\usepackage{empheq}

\newcommand{\be}{\begin{equation}}
\newcommand{\ee}{\end{equation}}
\newcommand{\bea}{\begin{eqnarray}}
\newcommand{\eea}{\end{eqnarray}}

\input{tcilatex}

\begin{document}

\title{Viable Gauge Choices in Cosmologies with Non-Linear Structures}

\author{Timothy Clifton}
\author{Christopher S. Gallagher}
\author{Sophia Goldberg}
\author{Karim A.~Malik}
\affiliation{School of Physics \& Astronomy, Queen Mary University of London, UK.}

\bibliographystyle{plain}

\date{\today}

\begin{abstract}
A variety of gauges are used in cosmological perturbation theory. These are often chosen in order to attribute physical properties to a particular choice of coordinates, or otherwise to simplify the form of the resultant equations. Calculations are then performed with the understanding that they could have been done in any gauge, and that transformations between different gauges can be made at will. We show that this logic can be extended to the domain of large density contrasts, where different types of perturbative expansion are required, but that the way in which gauges can be chosen in the presence of such structures is severely constrained. In particular, most gauges that are commonly considered in the cosmology literature are found to be unviable in the presence of non-linear structures. This includes {\it spatially flat} gauge, {\it synchronous} gauge, {\it comoving orthogonal} gauge, {\it total matter} gauge, {\it N-body} gauge, and the {\it uniform density} gauge. In contrast, we find that the longitudinal gauge and the Newtonian motion gauge are both viable choices in both standard cosmological perturbation theory, and in the post-Newtonian perturbative expansions that are required in order to model non-linear structures.

\end{abstract}


\maketitle

\section{Introduction}

Cosmological observations now span an enormous range of scales, from individual galaxies all the way up to the entire observable Universe. The theory of general relativity is widely held to govern the gravitational interaction, and therefore the dynamics of matter, over this entire range. However, within this theory the type of perturbative expansions used to model small fluctuations existing on very large scales are quite different from those which should be used on small scales. This paper investigates the mathematical structure of these different types of expansions, and uses the results to identify the choices of gauge that are viable on both large and small scales in cosmology.

Throughout this article we will make the assumption that space-time is everywhere close to a single Friedmann-Lema\^{i}tre-Robertson-Walker (FLRW) geometry, which can be written as
\begin{equation} \label{flrw}
ds^2 = \bar{g}_{\mu\nu} dx^{\mu} dx^{\nu} = a^2(\tau) \left( -d\tau^2 + dx^2+dy^2+dz^2 \right) \, ,
\end{equation}
where we have assumed spatial flatness (small amounts of spatial curvature can be included perturbatively). We also take the matter content of the space-time to be well modelled by a single perfect fluid, such that its stress-energy tensor can be written
\be \label{set}
T_{\mu \nu} = (\rho +p) u_{\mu} u_{\nu} + p \, g_{\mu \nu} \,,
\ee
where $\rho$ and $p$ are the energy density and isotropic pressure measured by an observer whose worldline is an integral curve of $u^{\mu}$. We have used letters from the Greek alphabet to denote spacetime indices, lower case Latin letters to denote space indices, and chosen units such that $c=1$. Treating the geometry and matter content of the Universe in this way are justified by their compatibility with a wide array of cosmological observables \cite{a1,a2,a3,a4,a5} (although it is not entirely without controversy \cite{b1,b2,b3}).

In order to include small inhomogeneities in the geometry of space-time one then writes the metric as
\be \label{dg}
g_{\mu \nu} = \bar{g}_{\mu \nu} + \delta g_{\mu \nu} \, ,
\ee
where the condition of smallness is enforced by the requirement $\delta g_{\mu \nu} \ll \bar{g}_{\mu \nu}$. While the coordinates used in writing the line-element in Eq.~(\ref{flrw}) are in some sense unique (up to spatial translations and rotations), the same cannot be said of the coordinates that one uses to express $\delta g_{\mu \nu}$: There is a ``gauge'' dependence, which can be viewed as a freedom to make an infinitessimal change in coordinates $x^{\mu} \rightarrow x^{\mu} + \xi^{\mu}$ (or, equivalently, as a change in mapping between the perturbed and unperturbed space-times). Without a single preferred set of coordinates one is forced to make a choice, and it is this that leads to the gauge problem in cosmology.

There are a wide array of gauges routinely used in cosmology to associate the coordinates with either preferred properties of the geometry or matter content. These may be, but are not limited to, coordinate-induced foliations that are spatially flat or orthogonal to the world-lines of observers, time coordinates that correspond to proper time of a class of observers, or choices that reduce the field equations to some desirable form. Choosing such a set of coordinates reduces the number of degrees of freedom that need to be solved for in a given physical problem, and removes the possibility of spurious gauge artefacts being introduced into the solutions. It also often allows the equations that describe that problem to be written in a simplified way. It is therefore highly desirable to understand the gauges that are possible, in any given situation.

We will investigate the viable gauge choices in cosmology for the following two weak-field expansions:
\begin{itemize}
\item[(i)] Cosmological perturbation theory,
\item[(ii)] Post-Newtonian theory,
\end{itemize}
which will both be discussed in detail in the next section. The former of these expansions is valid in a wide array of scenarios, as long as the density contrast and peculiar velocities of matter fields remain small, while the second is valid for arbitrarily large density contrasts, but only on small spatial scales. It is therefore the latter that should be used to describe the gravitational fields of the highly non-linear structures that exist at late times on scales $\lesssim 100 \, h^{-1}$Mpc. We find that most of the gauges that are commonly used in cosmology are not compatible with post-Newtonian theory, and therefore should not be used when modelling non-linear structures in the late universe. Exceptional cases are the longitudinal gauge and the Newtonian motion gauge, which are both valid in the presence of non-linear structures.

We begin in Section \ref{sectheory} with a general discussion of weak field expansions in cosmology, and their application in the form of cosmological perturbation theory and post-Newtonian gravity (suitably adapted for cosmology). In Section \ref{secgauge} we then discuss gauge transformations in these two types of theory, before progressing to a discussion of the commonly used gauges in cosmology in Section \ref{seccosmology}. Section \ref{nmsec} then contains a detailed analysis of the Newtonian motion gauge, where we find that this idea can be implemented in post-Newtonian gravity, and in cosmological perturbation theory. We then conclude in Section \ref{secdisc}. 

\section{Weak-Field Expansions in Cosmology}
\label{sectheory}

Weak field expansions are applicable in cosmology when the geometry of space-time is, in some sense, close to a known exact solution (usually FLRW). If this is the case, then we can write the metric of the space-time as that of the background universe plus a small perturbation, as in Eq. (\ref{dg}). If required, we can then irreducibly decompose the perturbations $\delta g_{\mu \nu}$ using Helmholtz theorem, so that they can be written as
\begin{eqnarray}
\delta g_{00} &=&  -2 a^2 \phi \label{dg00}\\
\delta g_{0i} &=&  a^2 (B_{,i} - S_i) \label{dg0i}\\
\delta g_{ij} &=& a^2 (-2 \psi \, \delta_{ij} + 2 E_{,ij} + 2 F_{(i,j)} + h_{ij}) \label{dgij}\, ,
\end{eqnarray}
where $S_i$ and $F_i$ are divergenceless vector field components, and where $h_{ij}$ is divergenceless and tracefree. In terms of these new quantities we can expand the 4-velocity of the fluid as
\be
u^{\mu} = \frac{1}{a} \left( 1- \phi + \frac{1}{2} v^2, v^i \right)  \, ,
\ee
where $v^i$ is the 3-velocity of the fluid. The Latin labels here correspond to spatial indices, and the factors of two and $a=a(\tau)$ are introduced for convenience only. Both cosmological perturbation theory and post-Newtonian theory are examples of weak field expansions, as they both typically have gravitational potentials of magnitude $\lesssim 10^{-4}$. We will consider these two expansions in what follows.

If we are attempting to model a particular physical situation with a weak-field expansion, then we need to identify which type of expansion(s) are applicable in that situation. Crucial factors in making such an assessment are the magnitude of the velocity and magnitude of the density contrast of matter in the system. In systems with small density contrasts we typically find $v \sim 10^{-4}$ (such that $\phi \sim v$), whereas in systems with highly non-linear matter we typically have $v \sim 10^{-2}$ (such that $\phi \sim v^2$). The former of these relationships follows directly from being in a situation in which a linearised version of the field equations are applicable, such that all deviations from the homogeneous and isotropic FLRW background are ``small'' (including the density contrast). On the other hand, the latter can be readily identified from the virial theorem, which is derived in the Newtonian limit of Einstein's equations, and which is thought to be a good approximations in most non-linear situations where $\phi$ and $v$ are both small.

When considering the applicability of different approaches to studying weak-field gravity, an important factor is the spatial extent of the domain under consideration. On spatial scales that are in some sense ``small'', and on which we have slowly moving matter, one can take a limit of a post-Minkowski expansion in order to find that the solutions to the field equations (which in general have null characteristic curves) can be well approximated by the solutions to Poisson equations (which have support on space-like hypersurfaces). This is what happens in post-Newtonian theory, and it formally changes the structure of the perturbative expansion by producing a hierarchy of PDEs in spatial variables only. In general, no such simplification can be made on horizon-sized scales, and so the system of equations we are required to solve remains a set of PDEs in both space and time. In this case post-Newtonian theory can no longer be applied.

\begin{figure}[t!]
\includegraphics[width=\columnwidth]{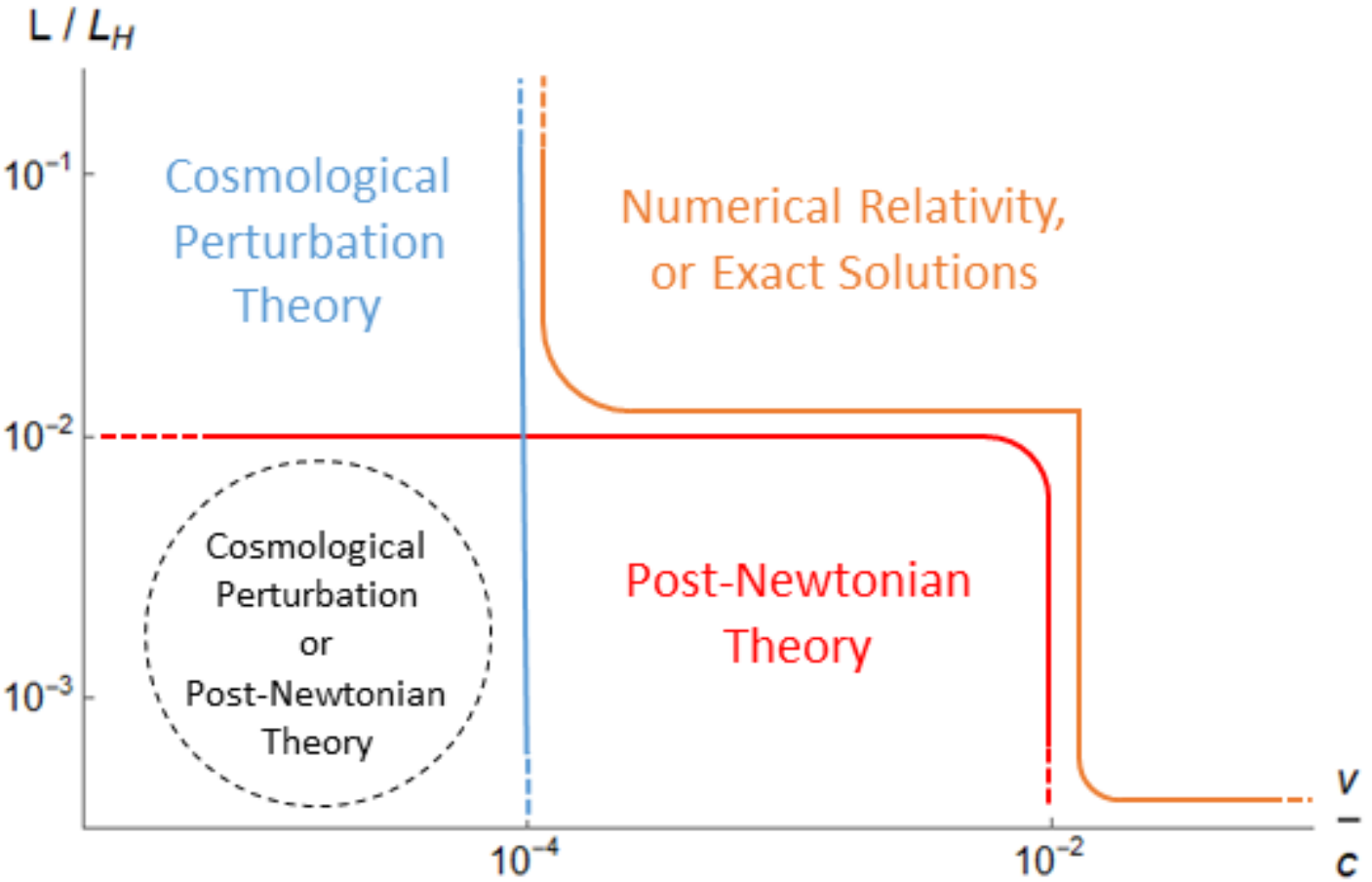}
\caption{Domains of applicability of cosmological perturbation theory (left of the blue line), and post-Newtonian theory (under the red line). Both formalisms are valid on small scales, when velocities (and hence density contrasts) are small. The characteristic spatial scale of a system is denoted $L$, and the Hubble scale is denoted $L_H$.}
\label{fig1}
\end{figure}

Fig. \ref{fig1} shows the domains of applicability of both cosmological perturbation theory and post-Newtonian theory. Cosmological perturbation theory can (in principle) by applied to {\it any} spatial scale, as long as the magnitude of all fluctuations from the FLRW background remain small. In the early universe this is widely thought to correspond to all spatial scales (unless primordial black holes are present), but in the late universe it means that scales on which the density contrast have become non-linear should not be expected to be well modelled using such an approach. On the other hand, for post-Newtonian theory to be applicable the range of spatial scales is restricted to those that are ``small'' compared to the cosmological horizon (this statement will be made more precise later on). However, when modelling systems on such scales using post-Newtonian theory, the density contrast is only restricted by the condition that it does not cause the velocity of matter to violate $v \ll 1$. This is valid all the way down to compact astrophysical objects, such as neutron stars and black holes.

In fact, it is of course already standard practise to apply linear cosmological perturbation theory to fluctuations on large scales (or on any scale at early times), and to use the equations of Newtonian gravity on the small scales where structures become non-linear at late times. These are the leading-order parts of cosmological perturbation theory and post-Newtonian theory, respectively. What is not currently done consistently in much of the literature is to use the next-to-leading order parts of each of these formalisms to simultaneously calculate relativistic gravitational effects from structures in each of their respective domains of applicability. Instead, cosmological perturbation theory is extrapolated down into the non-linear regime \cite{villa}, or post-Newtonian gravity is extended up to horizon-sized scales \cite{bruni}. A more comprehensive approach, in which both expansions are deployed together, was only recently introduced in Refs. \cite{sophia1, sophia2, kit1, kit2}. In this paper we will consider cosmological perturbation theory and post-Newtonian theory separately, in order to identify suitable gauge choices in each case.

\subsection{Cosmological Perturbation Theory}

Cosmological perturbation theory treats all perturbative objects on an equal footing, and expands all relevant equations order-by-order in every variable. This necessarily results in taking both the peculiar velocities and density contrasts to be ``small''. It is a strict application of the more general concept of perturbation theory to the case of almost-FLRW cosmologies. As such, every quantity is assigned an order-of-smallness using a parameter $\epsilon$. This includes both perturbations to the geometry, as well as perturbations to the background stress-energy tensor, such that in a perfect fluid-filled universe we have
\begin{equation} \label{epsilon}
\phi \sim \psi \sim v \sim \delta \sim S_i \sim B_{,i} \sim F_{i,j} \sim E_{,ij} \sim h_{ij} \sim \epsilon \ll 1 ,
\end{equation}
where $v$ is the peculiar velocity of the matter fields (as a fraction of the speed of light), $\delta$ is the density contrast, and all geometric variables are defined in Eqs. (\ref{dg00})-(\ref{dgij}). The reader will note that appropriate derivatives have been added to $B$, $E$ and $F_i$, in order to make them dimensionless. The comparison between spatial tensor, vector and scalar modes in this expression should be taken to mean that each component of the vectors and tensors are of the same order of magnitude as each of the scalars.

An important feature of cosmological perturbation theory is that, once the smallness of the quantities in Eq. (\ref{epsilon}) have been identified, all field equations and equations of motion can first be expanded to linear order, then quadratic order, and subsequently to all higher-orders in these variables. This gives, for the background part of the field equations, the usual Friedmann equations
\be \label{f1}
\mathcal{H}^2 = \frac{8 \pi G}{3} \bar{\rho} \, a^2 +\frac{\Lambda}{3} \, a^2
\ee
and
\be \label{f2}
\dot{\mathcal{H}} = -\frac{4 \pi G}{3} (\bar{\rho}+3 \bar{p}) \, a^2 +\frac{\Lambda}{3} \, a^2 \, ,
\ee
and energy conservation equation
\be \label{energybg}
\dot{\bar{\rho}} + 3 \mathcal{H} (\bar{\rho} + \bar{p} ) = 0 \, ,
\ee
where $\Lambda$ is the cosmological constant. For the scalar part of the field equations, to linear-order in $\epsilon$, we find the constraint equations
\bea \label{cpt1}
&&\nabla^2 \psi - 3 \mathcal{H} \left( \dot{\psi} + \mathcal{H} \phi \right) + \mathcal{H} \nabla^2 \sigma = 4 \pi G a^2 \delta {\rho} \\
&&\dot{\psi} + \mathcal{H} \phi = -4 \pi G a^2 (\bar{\rho} +\bar{p})\left( v + B \right) \, ,
\eea
where $v$ and $\sigma = \dot{E} -B$ are the scalar parts of the velocity and shear, respectively. We have also used dots to denote differentiation with respect to $\tau$, and written the Hubble rate as $\mathcal{H}=\dot{a}/a$. The perturbed conservation equations, which give the evolution equations for $\delta$ and $v$, can be written as
\bea
&&\hspace{-1cm} \delta \dot{\rho} + 3 \mathcal{H} (\delta \rho+ \delta p) = (\bar{\rho} + \bar{p}) \left[ 3 \dot{\psi} - \nabla^2 \left( v+ \dot{E} \right) \right] \label{cpt3} \\
&&\hspace{-1cm}\partial_{\tau}\left[ (\bar{\rho}+\bar{p}) (v+B) \right] + \delta p = - (\bar{\rho}+\bar{p}) \left[ \phi +4 \mathcal{H} (v+B)\right]  \label{cpt4}
\eea
where $\bar{p}$ and $\delta p$ are the background and perturbation to the isotropic pressure.
We note the evolution equation for $\sigma$ can be obtained from the linearised Einstein equations, and can be written as $\dot{\sigma} +2 \mathcal{H} \sigma - \phi + \psi = 0$.

The linear cosmological perturbation equations (\ref{cpt1})-(\ref{cpt4}) have a number of well known properties, which greatly aid one in finding and understanding their solutions. Firstly, it can be seen that one does not need to know anything about the divergenceless vector or tensor degrees of freedom in the metric, in order to write down a consistent set of equations that can be used to solve for the scalar parts of the gravitational field. This is a result of the decomposition theorem, which holds for all linear equations in cosmological perturbation theory (but does not hold at any higher order). A second property is that any derivatives acting on any quantities does not change its order of magnitude in the expansion, and neither does multiplication or division by the background quantities $\mathcal{H}$ (the Hubble rate) or $\bar{\rho}$ (the background energy density). This property is very important for the theory, as together with the fact that all of the quantities in Eq. (\ref{epsilon}) are dimensionless, it means that there is no limit to the spatial scales to which the theory is applied, unless that limit also happens to imply that one of the quantities in Eq. (\ref{epsilon}) is no longer small.

What we cannot do with cosmological perturbation theory, however, is expect it to provide accurate solutions when one of the quantities in Eq. (\ref{epsilon}) is no longer $\ll 1$. This is readily apparent from our attempts to model non-linear structures in the late universe. While we can use cosmological perturbation theory to extrapolate results into the mildly non-linear regime, it becomes highly problematic to try and use it find results when the density contrast becomes highly non-linear (which is exactly the reason cosmologists use Newtonian $N$-body simulations).  This problem arises because of the structure of the differential equations that result from applying cosmological perturbation theory, which at each order of the expansion produce a linear equation (or set of equations) in the new variables at that order, and which means that even the mild (quadratic) non-linearity that exists in the Newtonian equations of motion requires an infinite number of orders in perturbation theory in order to approach the true value (if the theory is convergent at all). 

In the highly non-linear regime it is much easier to follow the approach prescribed in Newtonian $N$-body simulations: To solve the linear Newton-Poisson equation, and then solve the non-linear Eulerian equations of motion in the resultant gravitational field. In this situation it is clear that the Newtonian limit of general relativity cannot be readily recovered from cosmological perturbation theory, and that it is not a limiting case of that approach. Let us now consider how the Newtonian limit can be realised in cosmologies with non-linear structures, as part of a consistent weak-field expansion.

\subsection{Post-Newtonian Theory}
\label{PNT}

In contrast to cosmological perturbation theory, post-Newtonian theory requires that perturbations to the various geometric and matter variables appear at different orders in the expansion of the field equations and equations of motion. It also requires that time-derivatives and space derivatives have different orders-of-magnitude associated with them. These departures from the usual approach used in standard cosmological perturbation theory increases its complexity, but has the very considerable advantage that it results in a theory that is valid in the presence of extremely large density contrasts (which are not formally part of this particular weak-field expansion at all). In this section we will outline how post-Newtonian theory can be used in cosmology.

An essential property of the post-Newtonian expansion is that it is a slow-motion, as well as weak-field, expansion. The condition of being slow-motion can be understood by identifying the length scales involved in the type of physical systems we wish to model \cite{poisson}. Let us start by identifying a characteristic time-scale for such a system, $t_{\rm c}$. This could correspond to the orbital period for two bodies in close proximity, or to the time-scale required for a large body (such as a cluster) to assemble itself. The gravitational field is known to propagate at the speed of light, $c$, which means that we can associate a characteristic length scale with the variations of our system:
\be
\lambda_{\rm c} = c \, \tau_{\rm c} \, .
\ee
Due to the high propagational speed of light, the value of $\lambda_{\rm c}$ is typically very large compared to the spatial scale of the system itself, which we denote as $r_{\rm c}$. Such a system will typically contain matter that has 3-velocities of magnitude $v_{\rm c} = {r_{\rm c}}/{\tau_{\rm c}}$, so it can immediately be seen that the slow-motion condition $v_{\rm c} \ll c$ is equivalent to the condition that $r_{\rm c} \ll \lambda_{\rm c}$. That is, systems that are considered slow motion should exist on scales that are much less than the characteristic length scale of the gravitational fields that are associated with them.

In cosmology we are interested in structures that grow over time-scales that are comparable to the age of the Universe (or less), so we have $\tau_{\rm c} \sim \mathcal{H}^{-1}$ and therefore $\lambda_{\rm c} \sim c \, \mathcal{H}^{-1}$ \cite{sophia1}. The characteristic length scale is therefore that of the observable universe, and the slow motion condition restricts us to considering systems that have a spatial extent that is much smaller than that scale. More precisely, we are limited by
\be
\frac{r_{\rm c}}{\lambda_c} \sim \frac{v_{\rm c}}{c} \sim 10^{-2} \, ,
\ee
where we have extracted a typical velocity from the virial relation $\phi \sim v^2 /c^2$, and used the empirical observation that we have at most $\phi \sim 10^{-4}$ for all systems of interest. For a Universe with $\lambda_{\rm c} \sim c \, \mathcal{H}^{-1} \sim 10^4 \,$Mpc this gives us $r_{\rm c} \sim 100 \,$Mpc. This shows that if we wish to apply a slow-motion condition, in the context of a weak-field expansion, then we should restrict the domain of applicability of such an approach to systems that have a spatial scale that are $\sim 100\,$Mpc (or less).

Let us now consider what the slow-motion condition implies for the form of the field equations, and their solutions. From Eq. (\ref{set}) we can immediately identify $T^{00} \simeq \rho \, c^2$, $T^{0i} \simeq \rho \, v^i c$ and $T^{ij} \simeq \rho \, v^iv^j +p \delta^{ij}$. These immediately imply
\be
\frac{T^{0i}}{T^{00}} \sim \frac{v_{\rm c}}{c} \qquad {\rm and} \qquad \frac{T^{ij}}{T^{00}} \sim \frac{v^2_{\rm c}}{c^2} \, ,
\ee
which from the field equations implies
\be
\frac{\delta g^{0i}}{\delta g^{00}} \sim \frac{v_{\rm c}}{c} \qquad {\rm and} \qquad {\delta g^{00}} \sim {\delta g^{ij}} \sim \frac{v^2_{\rm c}}{c^2} \, .
\ee
If we now recall that $\phi \sim v^2/c^2$, and choose units such that $r_{\rm c} \sim 1$, then we see that we can write
\be
\phi \sim \psi \sim F_{i,j} \sim E_{,ij} \sim h_{ij} \sim \rho \sim \eta^2
\ee
and 
\be
S_i \sim B_{,i} \sim \eta^3 \qquad {\rm and} \qquad p \sim \eta^4 \, ,
\ee
where we have introduced $\eta = v_{\rm c}/c \ll 1$ as the order of smallness in this expansion. The reader will note that different geometric and matter perturbations appear at different orders in the expansion, and that the density contrast does not have to be small at all.
 
From purely kinematic considerations we can immediately identify that the slow-motion criterion also has consequences for the order-of-smallness of quantities that contain derivatives. If the constituent parts of a system are moving slowly, then this immediately implies that the time-variation of state variables such as energy density and pressure will also be only slowly varying. This can be quantified in terms of $\eta$ as follows:
\be
\frac{\dot{\rho}}{\nabla \rho} \sim \frac{r_{\rm c}}{\tau_{\rm c}} \sim \eta \, ,
\ee
where we have used $\nabla$ to denote a spatial derivative, and where similar results should hold for derivatives of $v$ and $p$. If we again choose units such that $r_{\rm c} \sim 1$, then we find that time derivatives of matter variables add an extra order-of-smallness in $\eta$. 

Given that we expect gravitational perturbations on small scales to inherit the time dependence of the stress-energy tensor components that source them, this rule should extend to metric perturbations as well. This leads to the general rule
\be
\frac{\partial/\partial \tau}{\partial/\partial x}\sim \eta \, ,
\ee
i.e. that every time derivative adds an extra order-of-smallness, when acting on either matter fields or gravitational fields. This rule means that the field equations that would normally correspond to null wave equations can instead be written at leading-order as Poisson equations:
\be \label{wave} \nonumber
\square \delta g_{\mu \nu} \propto T_{\mu \nu} - \frac{1}{2} g_{\mu \nu} T \;\; \Rightarrow \;\; \nabla^2 \delta g_{\mu \nu} \propto T_{\mu \nu}- \frac{1}{2} g_{\mu \nu} T \, ,
\ee
where $\square = \bar{g}^{\mu \nu} \partial_{\mu} \partial_{\nu}$ and $\nabla^2 = \bar{g}^{ij} \partial_{i} \partial_{j}$ and $T=T^{\mu}_{\phantom{\mu} \mu}$.

\begin{figure}[tbh]
\includegraphics[width=0.8 \columnwidth]{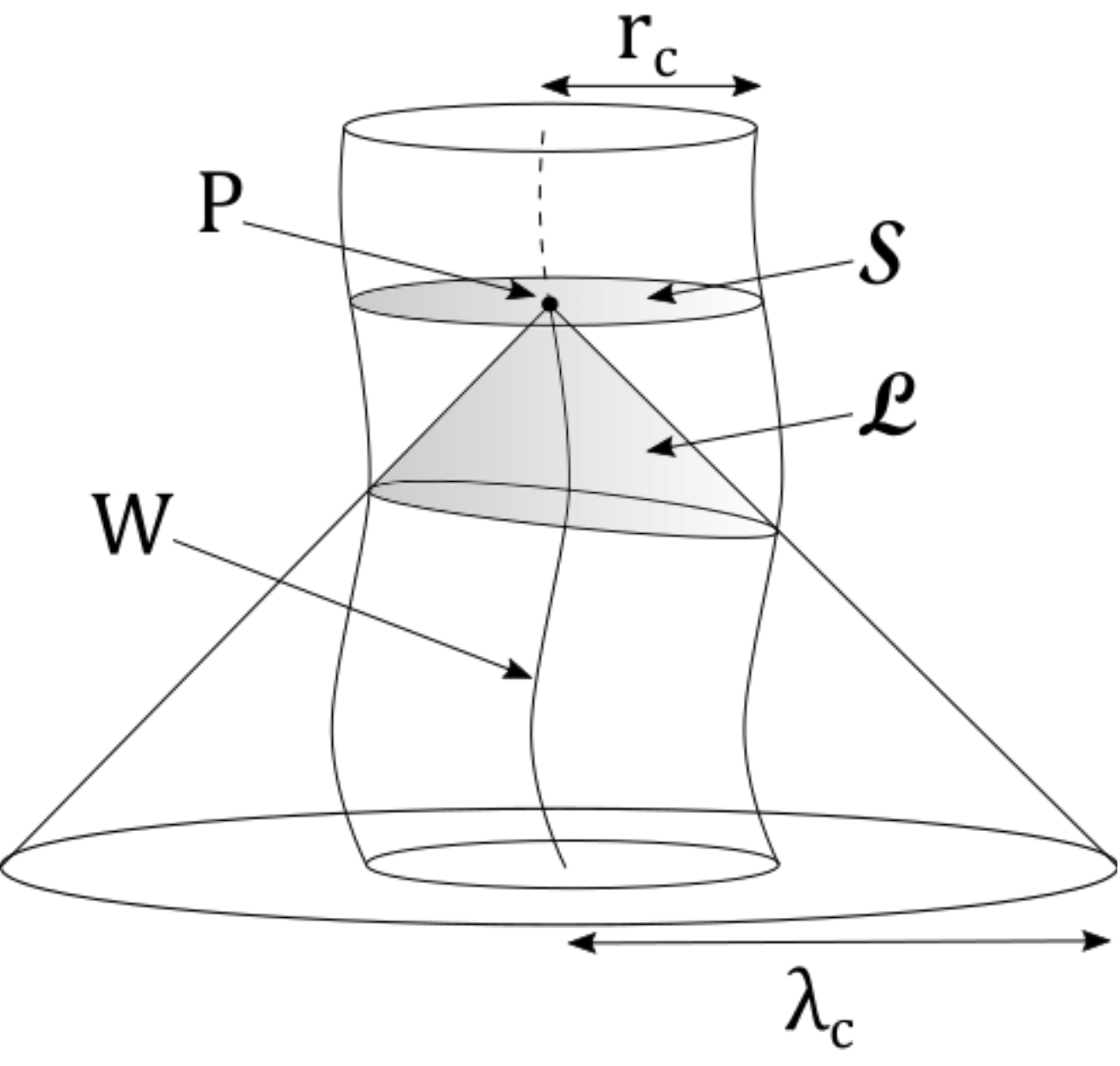}
\caption{The past lightcone $\mathcal{L}$ of a point $P$ following a worldline $W$. The support for the metric perturbations at $P$ can be approximated as being located on the space-like hypersurface $\mathcal{S}$, as long as $r_{\rm c} \ll \lambda_{\rm c}$.}
\label{fig2}
\end{figure}

The support for the integral that gives the function $\delta g_{\mu \nu} (t,{\bf x})$ in Eq. (\ref{wave}) should really be taken to be on the past light cone $\mathcal{L}$ of the point $P$ at position ${\bf x}$. This shows the causal nature of Einstein's theory, and the fact that gravitational interactions propagate at the speed of light. However, such an approach would be problematic to apply in cosmology, as the integral for the gravitational fields at each point in space would have its own distinct domain (i.e. its own past lightcone). A fortunate consequence of the slow-motion expansion is that on scales $r \lesssim r_{\rm c}$ we can approximate the past light cone of a point as being given by a space-like surface $\mathcal{S}$ of constant $\tau$ \cite{poisson}, as shown in Fig. \ref{fig2}. This is because the time taken for a null signal to go from one side of such a domain to the other is negligible compared to $\tau_{\rm c}$, and means that we can find solutions for $\delta g_{\mu \nu} (t,{\bf x})$ at some time $\tau$ by simply integrating over a suitable region of a hypersurface of constant $\tau$. The integrals for the gravitational field value at neighbouring points in space then have their support on overlapping domains, and the whole process of finding solutions is considerably simplified.

The lowest order fields, using the rules outlined above to $O(\eta^2)$, then give us the following constraint and evolution equations \cite{averaging}:
\be \label{pnfrw1}
\mathcal{H}^2 + \frac{2}{3} \nabla^2 \psi = \frac{8 \pi G}{3} \, \rho \, a^2 \, + \frac{\Lambda}{3} \, a^2+ O(\eta^4)
\ee
and
\be \label{pnfrw2}
\dot{\mathcal{H}} - \frac{1}{3} \nabla^2 \phi = -\frac{4 \pi G}{3} \, (\rho + 3 \bar{p}) \, a^2 \, + \frac{\Lambda}{3} \,a^2+ O(\eta^4) \, ,
\ee
where $\mathcal{H} \sim \tau_{\rm c}^{-1} \sim \eta$ and $\dot{\mathcal{H}} \sim \tau_{\rm c}^{-2} \sim \eta^2$ (in units such that $r_{\rm c} \sim 1$). These equations are a combination of the Hubble equations and the Newton-Poisson equations for $\phi$ and $\psi$, which both occur at the same order in this expansion. Within a region of space $\mathcal{S}$, of scale $r \lesssim r_{\rm c}$, they can be transformed to the usual Newtonian equations through a suitable choice of coordinates. It is also known that many such regions can be patched together to form a cosmology described by a line-element that is close to a single global FLRW solution \cite{sanghai}, as in Eq. (\ref{flrw}).

If we integrate Eqs. (\ref{pnfrw1})-(\ref{pnfrw2}) over $\mathcal{S}$, and divide by the spatial volume of that region, we recover the standard Friedmann equations (\ref{f1})-(\ref{f2}), as well as the Newton-Poisson equations in an expanding background,
\be
\nabla^2 \phi = \nabla^2 \psi = 4 \pi G \, \delta \rho \, a^2 \; ,
\ee
as long as we choose the boundary condition
\be \label{bc}
\int_{\partial \mathcal{S}} {\bf \nabla} \phi \cdot d{\bf S} = 4 \pi G \, \langle \delta \rho \rangle \, a^2 \, ,
\ee
where we have written $\rho = \bar{\rho}+ \delta \rho$, and where $\langle \delta \rho \rangle$ is the volume averaged value of $\delta \rho$ in the region $\mathcal{S}$. It is important to note that there is no assumption made about the relative sizes of $\bar{\rho}$ and $\delta \rho$ here; the post-Newtonian expansion is specifically constructed to allow for large density contrasts to be consistently modelled, and this means that $\delta \rho/\bar{\rho}$ is allowed to be much larger than one without signalling any breakdown in the weak-field expansion.

The left-hand side of Eq. (\ref{bc}) can be set to zero if one chooses $\mathcal{S}$ to have periodic boundary conditions, which also sets the right-hand side to zero (as the average of this spatial domain would automatically be equal to the global average of the cosmology). In general, it seems conceivable that Eq. (\ref{bc}) may not be satisfied. If this is so, then one should expect strong cosmological back-reaction, and a violation of our initial ansatz of a perturbed FLRW space-time, but we will not consider this further here.

It is noteworthy that the Friedmann equations and the Newton-Poisson equations occur at the same order of magnitude in this expansion. This shows the well known fact that post-Newtonian expansions are {\it not} (strictly speaking) a direct application of perturbation theory, a fact that is already obvious from the leading-order conservation equations:
\be \label{energy}
\dot{\rho} +3 \mathcal{H} \rho + \partial_i \left( \rho v^i \right) = 0\, ,
\ee
and
\be \label{momentum}
\rho \dot{v}_j +\rho v^i \partial_i v_j + \rho \mathcal{H}  v_j = -\rho \partial_j \phi - \partial_j p \, ,
\ee
which are the standard equations of Newtonian gravity on an expanding background. These equations are clearly non-linear, and therefore cannot be considered as being the result of an application of perturbation theory, in the strict sense outlined in the previous section. Nevertheless, they are well-defined, and the post-Newtonian expansion itself constitutes a well-defined expansion of the field and conservation equations, which has been extensively applied in other areas of gravitational physics.

All equations in this section, as well as higher-order equations, can be obtained by direct coordinate transformation from their form in the post-Minkowski approach \cite{sanghai}. Their existence shows the direct correspondence (through an isomorphism) of the expansion about a Friedmann space that we have just outlined, and the extremely well studied expansions that are usually performed around Minkowski. They can be used to further justify the order of magnitude we have associated with the various quantities we have required, as well as understanding some of the features that have become apparent. Firstly, the applicability to scales $r\ll \mathcal{H}^{-1}$ can be seen to correspond directly to the requirement that $v \ll c$. Secondly, the mixing of Friedmann and Newton-Poisson equations can be shown to be a result of the leading-order part of the cosmological expansion arising from the motion of particles under the influence of Newtonian gravitational fields in the perturbed Minkowski approach. We refer the reader to Refs. \cite{averaging, sanghai} for more details of these observations.

\section{Gauge Transformations}
\label{secgauge}

A gauge transformation is a transformation (or set of transformations) that preserve the structure of a theory, and that can be used to remove (or fix) redundant degrees of freedom. In general relativity the term {\it gauge} is sometimes used to describe the invariance of the theory under general coordinate transformations. When applied to a weak-field expansion of the field equations, however, it refers to the set of transformations that leave the theory that results from that expansion unchanged. In such cases the transformations must be small (in the sense of the expansion), and must provide a map from the full set of perturbed quantities into itself (see e.g. \cite{mm})

A gauge transformation can said to be either {\it active} or {\it passive}. We will use the former of these approaches, which changes the point in space-time that a given set of coordinate values identifies. The action of such a transformation on a tensor field ${\bf T}$ can be written in the form
\be \label{gt}
{\bf T} \rightarrow \tilde{{\bf T}} = e^{\mathcal{L}_{{\bf \xi}}} {\bf T} \, ,
\ee
where $\mathcal{L}_{{\bf \xi}}$ is the Lie derivative with respect to the gauge generator ${\bf \xi}$, and where a tilde denotes the field ${\bf T}$ after the transformation. The exponential map is used here to ensure that the group structure of the diffeomorphisms associated with the transformations is preserved. 

Treating the coordinates on the manifold as a set of four scalar fields, the gauge transformation in Eq. (\ref{gt}) can be seen to be equivalent to
\be
x^{\mu} (p) \rightarrow x^{\mu} (q) = e^{\xi^{\alpha} \partial_{\alpha} \vert_{p}} x^{\mu} (p) \, ,
\ee
where $x^{\mu}=x^{\mu}(p)$ on the right-hand side is evaluated at some point $p$, while $x^{\mu}(q)$ is evaluated at the point $q$ located along the flow of the gauge generator field ${\bf \xi}$ from $p$. Geometrically, one can think of this as a transformation in the map used to identify a point in the background space-time with a point in the perturbed space-time.
Let us now focus on the set of active gauge transformations that are possible in both cosmological perturbation theory and post-Newtonian theory.

\subsection{Cosmological Perturbation Theory}

If we apply the transformation in Eq. (\ref{gt}) to the metric we obtain:
\be \label{gtrans}
\tilde{g}_{\mu \nu} = g_{\mu \nu} + \mathcal{L}_{\xi} \, g_{\mu \nu} + \frac{1}{2} \mathcal{L}^2_{\xi} \, g_{\mu \nu} +  \, \dots \, ,
\ee
which gives the various components of the metric transforming as
\bea
g_{00} &\rightarrow& g_{00} + \xi^{\mu} \partial_{\mu} g_{00} + 2 g_{0\mu}\, {\dot{\xi}^{\mu}} + \dots \label{g00trans}\\
g_{0i} &\rightarrow& g_{0i} + \xi^{\mu} \partial_{\mu} g_{0i} + g_{0\mu}\, {\xi^{\mu}_{,i}} + g_{i\mu}\, {\dot{\xi}^{\mu}} + \dots \label{g0itrans} \\
g_{ij} &\rightarrow& g_{ij} + \xi^{\mu} \partial_{\mu} g_{ij} + 2 g_{\mu (i}\, {\xi^{\mu}}_{,j)} +\dots \, \label{gijtrans} 
\eea
where the ellipses in these expression denote terms that are quadratic or higher-order in the gauge generator, $\xi^{\mu}$. It can be seen immediately from these equations that every component of the gauge generators must be of the same order-of-magnitude in the perturbative expansion as the metric perturbations, i.e. that
\be
\xi^{\mu} \sim \epsilon \, .
\ee
If $\xi^{\mu}$ were larger than this, then the metric (after the gauge transformation), could no longer be written as perturbed FLRW.

Using this information, and the decomposition given in
Eqs.~(\ref{dg00})-(\ref{dgij}), we then find the standard set of gauge
transformations:
\bea
\phi &\rightarrow& \phi + \mathcal{H} \, \xi^0 + {\dot{\xi}^0} \nonumber \\
B &\rightarrow& B + \dot{\zeta} - \xi^0 \nonumber\\
S_i &\rightarrow& S_i - \dot{\zeta}_i\nonumber\\
\psi &\rightarrow& \psi - \mathcal{H} \, \xi^0 \nonumber\\
E &\rightarrow& E + \zeta \nonumber\\
F_i &\rightarrow& F_i + \zeta_i \nonumber\\
h_{ij} &\rightarrow& h_{ij} \, ,
\label{cosmo_pert_gauge_trans1}
\eea
where we have decomposed the spatial part of the gauge generator, such that $\xi_i = \zeta_{,i} + \zeta_i$, where $\zeta^i$ is divergenceless. It can be seen that {\it all} metric perturbations transform under a general gauge transformation, with the notable exception of $h_{ij}$ (at linear order).

Similarly, we can calculate how the components of the stress-energy tensor transform under a gauge transformation with $\xi^{\mu} \sim \epsilon$. For a perfect fluid, these components can be written
\bea
T^{0}_{\phantom{0} 0} &=& - (\bar{\rho} + \delta \rho)\\
T^{0}_{\phantom{0} i} &=& (\bar{\rho} +\bar{p}) (v_i + B_{,i} -S_i)\\
T^i_{\phantom{i} j} &=& (\bar{p} + \delta {p}) \delta^i_{\phantom{i}j}
\eea
and under the transformation (\ref{gt}) therefore give
\bea
\delta &\rightarrow& \delta + \xi^0 \, \frac{\dot{\bar{\rho}}}{\bar{\rho}} \nonumber\\
\delta p &\rightarrow& \delta p  + \xi^0 \, \dot{\bar{p}}\nonumber\\[5pt]
v^i &\rightarrow& v^i - \dot{\xi}^{i} \, .
\label{cosmo_pert_gauge_trans2}
\eea
It is again apparent that all perturbed quantities transform under a general gauge transformation, in the matter sector as well as the gravitational sector.

The gauge transformations given above are very well known in perturbation theory. What is less well known in cosmology are the transformation properties of post-Newtonian variables under the most general possible gauge transformation. We will spell this out below.

\subsection{Post-Newtonian Theory}
\label{pngaugesec}

Let us now consider linear gauge transformations of the metric, as given to linear order in $\xi^{\mu}$ by Eqs. (\ref{gtrans})-(\ref{gijtrans}). These equations take exactly the same form in post-Newtonian theory as they do in cosmological perturbation theory, as so far they have only assumed that $\xi^{\mu}$ is small, and that terms quadratic or higher can therefore be neglected to leading order.

Let us now consider the size of each of the terms that results from the gauge transformation in Eq. (\ref{gtrans}). Starting with the $ij$-component of the metric, we can see that the transformation (\ref{gijtrans}) has terms of magnitude
\be
g_{ij} \rightarrow g_{ij} + O(\xi^i) + O(\eta \, \xi^0) \, ,
\ee
where we use $O(x)$ to mean terms of order $x$ or smaller in the post-Newtonian expansion. In deriving this expression we have used the rules for the order-of-magnitude of each of the components of the metric, and the relative size of their derivatives, as outlined in Section \ref{PNT}. We have also taken the components of the gauge generator $\xi^{\mu}$ to obey the same rules with respect to derivative operators (i.e. that time derivatives of these objects are small compared to space derivatives).

Performing the same analysis for the $0i$-component of the metric we find
\be
g_{0i} \rightarrow g_{0i} + O(\eta \, \xi^i) + O(\xi^0) \, .
\ee
In order for the gauge transformed $ij$ and $0i$-components of the metric to be no larger than $\eta^2$ and $\eta^3$, respectively, we can see that we must have 
\be \label{xiorders}
\xi^i \sim \eta^2 \qquad {\rm and} \qquad \xi^0 \sim \eta^3 \, .
\ee
If the former of these conditions was violated, and the magnitude of $\xi^i$ were allowed to be larger than $\eta^2$, then it can be seen that the gauge transformed $ij$-components of the metric would have terms larger than $\eta^2$. This would mean that they would be larger than allowed in the post-Newtonian expansion of the metric, and the transformation would not be part of the gauge group of the theory. Similarly, if the magnitude of $\xi^0$ were allowed to be any larger than $\eta^3$ then the gauge transformed $0i$-components of the metric would contain parts that were larger than $\eta^3$, which is also forbidden for the same reason.

The appearance of different components of the gauge generators with different orders of magnitude is entirely absent from the approach to cosmological perturbation theory, but is entirely consistent with the post-Newtonian approach to gravity \cite{sophia1,sophia2}. Indeed, we have already seen that the perturbations to different components of the metric can have leading-order parts with different orders of magnitude. The consequences of this, however, do produce results that are unexpected from the perspective of the standard approach to perturbations in cosmology, as the different orders of magnitude of the different components of the gauge generators significantly alter the possible transformations of the metric. 

For example, if we investigate the leading-order terms that are generated in the transformation of the $00$-component of the metric we now find
\be
g_{00} \rightarrow g_{00} + O(\eta^2 \, \xi^i) + O(\eta \, \xi^0) \, ,
\ee
which, using Eq. (\ref{xiorders}), can be seen to be equivalent to
\be
g_{00} \rightarrow g_{00} + O(\eta^4) \, .
\ee
This means that the leading-order perturbation to the $00$-component of the metric, which exists at order $\eta^2$, is entirely unchanged by the gauge transformations that this theory admits, and that only sub-leading terms are affected. This result severely limits what can be done with gauge transformations in the presence of non-linear structures when using post-Newtonian expansions.

Having identified the orders of magnitude of the leading-order parts of the gauge generators, we can now find the leading-order parts of the gauge transformations of each of the degrees of freedom in the metric. These are given by
\bea
g_{00} &\rightarrow& g_{00} + \xi^{0} \dot{g}_{00}+ \xi^{i} g_{00,i} + 2 g_{00}\, {\dot{\xi}^{0}} + O(\eta^5) \\
g_{0i} &\rightarrow& g_{0i} + g_{00}\, {\xi^{0}_{,i}} + g_{ij}\, {\dot{\xi}^{j}} + O(\eta^4) \\
g_{ij} &\rightarrow& g_{ij} + 2 g_{k (i}\, {\xi^{k}}_{,j)} + O(\eta^3) \, ,
\eea
which gives
\bea
\phi &\rightarrow& \phi + \mathcal{H} \, \xi^0 + {\dot{\xi}^0} + \phi_{,i} \xi^i \\
B &\rightarrow& B + \dot{\zeta} - \xi^0  \\
S_i &\rightarrow& S_i - \dot{\zeta}_i  \\
\psi &\rightarrow& \psi  \\
E &\rightarrow& E + \zeta  \label{pnEtrans}\\
F_i &\rightarrow& F_i + \zeta_i  \label{pnFtrans}\\
h_{ij} &\rightarrow& h_{ij}   \, ,
\eea
where we have kept terms up to order $\eta^4$ in $\phi$, as this is the order required to obtain the post-Newtonian equations of motion for massive test particles. The reader will note that as well as the leading-order part of $\phi$ (at order $\eta^2$) being gauge invariant, the same can also be said of the leading-order parts of $\psi$ and $h_{ij}$. Only the last of these was invariant under general gauge transformations in cosmological perturbation theory.

Expanding the stress-energy tensor in the parameter $\eta$ we find
\begin{eqnarray}
T^{0}_{\phantom{0} 0} &=& - \rho (1 + v^2) + O( \eta^5) \\
T^{0}_{\phantom{0} i} &=& \rho v_i + O( \eta^4) \\
T^i_{\phantom{i} j} &=& \delta^i_{\phantom{i}j} p + \rho v^i v_j + O( \eta^5) \, ,
\end{eqnarray}
which under the gauge transformation (\ref{gt}) gives
\bea
\mu &\rightarrow& \mu\\
\Pi &\rightarrow& \Pi + \xi^i (\ln \mu)_{,i}\\
p &\rightarrow& p\\
v^i &\rightarrow& v^i \, ,
\eea
where we have written $\rho = \mu (1+ \Pi)$, such that $\mu \sim \eta^2$ is the rest-mass density and $\Pi \sim \eta^2$ is the specific energy density. All lowest-order parts of the matter variables can be seen to transform trivially, with an additional term at order $\eta^4$ appearing in the transformation of $\Pi$.

\section{Standard Gauge Choices in Cosmology}
\label{seccosmology}

Choosing a gauge is often essential in cosmology. However, the majority of gauges that are frequently used in the literature are not viable choices in the presence of non-linear structures modelled by post-Newtonian theory.  In this section we will review some of the ``popular'' gauges used in cosmological perturbation theory (see e.g.~Ref.~\cite{Malik:2008im} for details). 
These gauges are usually specified by assigning a particular set of variables to zero, either in the gravitational sector or the matter
sector (or a mixture of both). In each case we will also comment on the whether such a gauge can be achieved in the post-Newtonian expansion.

\subsection{Spatially Flat Gauge}

The spatially flat gauge is defined by the choice
\be
\psi=E=F_i=0\,,
\ee
which leaves the induced 3-metric on spatial hypersurfaces unperturbed (in the absence of tensor perturbations). This gauge is often used for the calculation of observables during inflation.

It can be seen from Eq. (\ref{cosmo_pert_gauge_trans1}) that this gauge can be readily achieved in cosmological perturbation theory by choosing $\xi^0=\psi/\mathcal{H}$, and $\xi_i = -E_{,i} -F_i$. On the other hand, in the post-Newtonian theory $\psi$ is gauge-invariant, and so this gauge is impossible to realise (though it is possible to set $E$ and $F_i$ to zero).

\subsection{Synchronous Gauge}

Synchronous gauge is defined by setting
\be
\phi=B=S_i=0\,.
\ee
This gauge is popular for numerical studies, but does not uniquely define the time-slicing (this can be fixed by choosing an additional gauge condition, for example that the perturbed dark matter 3-velocity vanishes).  In this gauge it can be seen that the time coordinate corresponds to the proper time of comoving observers at fixed spatial coordinates. Synchronous gauge is routinely used in a wide variety of cosmological calculations, and is the default gauge for CMBFAST \cite{cmbfast} and CAMB \cite{camb}.

This gauge is obtained within cosmological perturbation theory by solving the differential equations $\dot{\xi}^0 + \mathcal{H} \xi^0 = - \phi$ and $\dot{\xi}_i - \xi^0 = -B_{,i} +S_i$. However, it cannot be achieved in post-Newtonian theory as in this case $\phi$ is gauge invariant at leading order (though $B$ and $S_i$ are not).

\subsection{Comoving Orthogonal Gauge}

The comoving orthogonal gauge is defined by the gauge conditions
\bea
v_i=0\qquad {\rm and} \qquad B_{,i}=S_i\,,
\eea
which states that the fluid 3-velocity and 3-momentum vanish. In this gauge the constant time hypersurfaces are orthogonal to the fluid 4-velocity. In cosmological perturbation theory this gauge choice requires $\dot{\xi}^i=v^i$ and $\xi^0_{,i} = \dot{\xi}_i$. Once more, this gauge choice cannot be realised in post-Newtonian theory, this time because $v^i$ is gauge invariant at leading order (though $B_{,i}$ and $S_i$ are not, and could be set equal).

\subsection{Total Matter Gauge}

The total matter gauge is related to the comoving orthogonal gauge. It
has the gauge conditions
\be
v+B=0\qquad {\rm and} \qquad E=0=F_i\,.
\ee
Evaluating the density contrast in the total matter gauge, and the metric potential $\phi$ in the longitudinal gauge, allows one to write the cosmological perturbation equations in the form of a Poisson equation, equivalent to its Newtonian counterpart \cite{Wands:2009ex,Hidalgo:2013mba}. This gauge can be realised in cosmological perturbation theory by choosing $\xi^0 = v+B$ and $\xi_i = -E_{,i} - F_i$. It cannot be realised in post-Newtonian theory as the condition $v+B=0$ has parts at order $\eta$ and $\eta^3$, the former of which cannot be satisfied as it corresponds to $v=0$, and $v$ is gauge invariant (though the other conditions are again possible).

\subsection{Uniform Density Gauge}

In the uniform density gauge we use the density perturbation, or equivalently the density contrast, to specify the temporal gauge
condition
\be
\delta\rho=0\,.
\ee
To fix the spatial gauge we can choose, for example, $E=0=F_i$. In cosmological perturbation theory this choice of specification of the temporal gauge can be written as $\xi^0=-\delta \bar{\rho}/\dot{\bar{\rho}}$, but such a condition is impossible to implement in the post-Newtonian approach as $\mu$ is gauge invariant in this set-up.

\subsection{N-body gauge}

The $N$-body gauge is formulated in a situation where
\begin{equation}
v+B=0 \, ,
\end{equation}
as in the total matter gauge, above. The remaining gauge freedoms are then used to set the so-called ``counting density'' associated with $N$ bodies equal to the leading-order part of the energy density. This condition requires that the scalar deformation of the spatial volume is set to zero, which can be written as \cite{nbody}
\begin{equation} \label{nbg}
\psi +\frac{1}{3} \nabla^2 E = 0 \;.
\end{equation}
This can be achieved in cosmological perturbation theory by taking $\xi^0 = v+B$ and setting the spatial gauge using the solution of $\nabla^2 \zeta = 3 \mathcal{H} (v+B) - \nabla^2 E - 3 \psi$. Now, while $v+B=0$ still cannot be realised in post-Newtonian gravity, the condition given in Eq. (\ref{nbg}) is achieved by taking $\nabla^2 \zeta = -\nabla^2 E - 3 \psi$. It may therefore be possible to develop new variants of the $N$-body gauge with alternative specification of the temporal gauge condition, such as the N-boisson gauge \cite{nb1,nb2}.

\subsection{Longitudinal Gauge}

Longitudinal gauge (also referred to as {\it conformal Newtonian}, or as part of {\it Poisson} gauge) is defined by the scalar gauge conditions
\be
B=E=0\,.
\ee
As the scalar shear is given by $\sigma=\dot E-B$, this gauge is also known as ``zero-shear'' gauge (the spatial hypersurfaces have vanishing shear). This gauge conditions give a diagonal metric tensor for the scalar perturbations, which considerably simplifies calculations.  If there is no anisotropic stress, the field equations in this gauge give $\psi=\phi$, which allows one to write the governing field equations from cosmological perturbation theory in a form that is very close to the Newtonian equation.

This is the {\it only} standard gauge choice we have found in the cosmology literature that can be fully specified in both cosmological perturbation theory and post-Newtonian theory. This is achieved in both cases by taking $\xi^0 = B+\dot{E}$ and $\zeta = -E$. This gauge choice therefore appears to be particularly valuable if one wishes to perform calculations in both the linear and non-linear regimes of cosmology, and to find results in each case that can be consistently related to one another.

\section{Newtonian Motion Gauge}
\label{nmsec}

The {\it Newtonian motion gauge} was recently introduced by Fidler {\it et al} in Ref. \cite{nm1}, and further developed in Ref. \cite{nm2}. It is based on the idea of fixing a gauge such that the gravitational field equation and equations of motion of test particles take the same form that they do in the Newtonian problem, i.e. such that 
\bea \label{energynm}
&&\dot{\tilde{\mu}} +3 \mathcal{H} \tilde{\mu} + \partial_i \left(
\tilde{\mu} \tilde{v}^i \right) = 0\\ \label{momentumnm} &&\tilde{\mu}
\dot{\tilde{v}}_j +\tilde{\mu} \tilde{v}^i \partial_i \tilde{v}_j +
\tilde{\mu} \mathcal{H} \tilde{v}_j = -\tilde{\mu} \partial_j
\tilde{U} - \partial_j \tilde{p} \, , 
\eea 
where $\tilde{U}$ must satisfy an equation of the form \be \nabla^2 \tilde{U} = 4\pi \, \delta\tilde{ \mu} \, a^2 \, .  \ee The variables $\tilde{v}^i$,
$\tilde{\mu}$, $\tilde{U}$ and $\tilde{p}$ can be seen to satisfy equations of exactly the same form as the Newtonian equations (\ref{energy}) and (\ref{momentum}), but are not themselves the Newtonian variables. Instead, they should be thought of as variables that are constructed from objects that are defined in the corresponding relativistic problem.

This is a very interesting idea, as almost all N-body simulations are based on the equations that result from considering Newtonian physics on an expanding background. The Newtonian motion gauge therefore allows Newtonian N-body simulations to be interpreted in a relativistic context, and therefore for relativistic gravitational effects to be extracted from non-relativistic simulations. This is achieved by deforming the coordinate system (using gauge transformations) such that the coordinate positions of particles are the same as those that would appear in the Newtonian problem. Here we will investigate this idea in the context of cosmological perturbation theory and post-Newtonian theory.

\subsection{Cosmological Perturbation Theory}

It is clear that the non-linear equations (\ref{energynm}) and (\ref{momentumnm}) will not be able to be satisfied by the linearised equations of first-order cosmological perturbation theory. In order to establish whether or not this gauge is viable in such an approach, we therefore propose to expand Eqs. (\ref{energynm})-(\ref{momentumnm}) perturbatively, and see whether or not the equations of cosmological perturbation theory can be manipulated into the form of the equations that result.

We start by writing
\bea
\tilde{\mu} &=& \tilde{\bar{\mu}} + \delta \tilde{\mu} + O(\epsilon^2) \\
\tilde{v}^i &=& \delta \tilde{v}^i + O(\epsilon^2)  \, .
\eea
To background order we find that Eq. (\ref{energynm}) can be written as
\be
\dot{\tilde{\bar{\mu}}} +3 \mathcal{H} \, \tilde{\bar{\mu}} = 0 \, ,
\ee
which is clearly of the same form as the energy conservation equation (\ref{energybg}), as long as $\bar{p}=0$, whilst the momentum conservation equation (\ref{momentumnm}) is automatically satisfied. We therefore have $\tilde{\bar{\mu}} = \bar{\rho}$, and the requirement $\bar{p}=0$ (i.e. that we consider dust, at the level of the background).

Next, we can study the perturbed equations at first order. For Eqs. (\ref{energynm})-(\ref{momentumnm}) this gives
\bea
\delta \dot{\tilde{\mu}} + 3 \mathcal{H}\, \delta \tilde{\mu} + \tilde{\bar{\mu}}\, \delta\tilde{v}^i_{\phantom{i},i} &=& 0 \label{energynmlin} \\
\tilde{\bar{\mu}} \, \delta \dot{\tilde{v}}^j + \tilde{\bar{\mu}} \mathcal{H} \, \delta \tilde{v}^j &=& - \tilde{\bar{\mu}} \, \tilde{U}_{,j} - \delta {\tilde p}_{,j} \, . \label{momentumnmlin}
\eea
If we now consider the equation of energy conservation at first order in cosmological perturbation theory (\ref{cpt3}), then we see that if we choose $\delta \tilde{\mu} = \delta \rho - 3 \bar{\rho} \psi +\bar{\rho} \nabla^2 E$ and $\delta \tilde{v}^i = v^i $ then we can write this equation in the form of the linearised Newtonian equation (\ref{energynmlin}). This gives us
\be \label{mucpt}
\tilde{\mu} = \bar{\rho} + \delta \rho - 3 \bar{\rho} \, \psi  +\bar{\rho} \, \nabla^2 E + O(\epsilon^2)
\ee
and
\be \label{vcpt}
\tilde{v} = v + O(\epsilon^2) \, ,
\ee
where $\tilde{v}$ and $v$ are the scalar parts of $\tilde{v}^i$ and $v^i$, respectively. For this correspondence to follow we also require $\delta p=0$ (i.e. that the requirement to consider dust is extended to linear order).

The combination of variables used to construct $\tilde{\mu}$ and $\tilde{v}$ in Eqs. (\ref{mucpt}) and (\ref{vcpt}) have not yet required any choice of gauge. Let us now consider the linearised momentum conservation equation (\ref{momentumnmlin}) that these variables must satisfy. Substituting in from Eq. (\ref{vcpt}), and taking $\delta \tilde{p}=0$, we find that the following equation must be satisfied:
\be \label{dvnm}
\dot{B} + \mathcal{H} \, B = \tilde{U} - \phi \, ,
\ee
where $\tilde U$ must now satisfy
\be \label{Unmcpt}
\nabla^2 \tilde{U} = 4\pi \, a^2 \,\bar{\rho} \, ( \delta - 3 \psi + \nabla^2 E ) \, .
\ee
This derivation of this equation has used the Euler equation (\ref{cpt4}) from cosmological perturbation theory in order to eliminate $\dot{v}$, and can be seen to be equivalent to Eq. (4.5) of Ref. \cite{nm2} (though without specifying any restriction on the time gauge).

Further manipulation, using the linear equations from cosmological perturbation theory with $\delta p=0$, we find that Eq. (\ref{dvnm}) can be re-written as 
\be \label{dvnm2}
\boxed{
\ddot{E} +\mathcal{H} \dot{E} - 4 \pi \, a^2 \, \bar{\rho} E = 3 \, \bar{\rho} \, \Phi_\mathcal{R} 
} \, ,
\ee
where 
\be
\Phi_\mathcal{R}  = - a^2(\tau) \int \frac{\mathcal{R}'}{\vert {\bf x} - {\bf x'} \vert} d^3 x'
\ee
and where $\mathcal{R}= \psi - \mathcal{H}(v+B)$ is the curvature perturbation in comoving orthogonal gauge (a well-known gauge invariant quantity, frequently used in cosmology). The boxed equation (\ref{dvnm2}) needs to be satisfied if the Newtonian motion gauge is to be realised in cosmological perturbation theory.

At this point it seems worth pointing out that the choices for the effective Newtonian variables made in Eqs. (\ref{mucpt}) and (\ref{vcpt}) are not unique, though they did lead to a viable application of the idea of a Newtonian motion gauge. We explore an alternative choice in Appendix \ref{app0}. There also remain gauge freedoms in the generators $\xi^0$ and $\zeta^i$, which can be set to any convenient values without affecting the property that the equations of motion can be written in a Newtonian form.

\subsection{Post-Newtonian Theory}

The lowest-order parts of $T^{\mu \nu}_{\phantom{\mu \nu} ; \nu}=0$ very obviously give equations that are in the form of the Newtonian equations of motion in post-Newtonian theory, as this is exactly how the Newtonian limit is derived in the context of relativistic gravity. The challenge in this case is therefore to put the equations of motion at first post-Newtonian order into the form of the Newtonian equations.

The relativistic field equations and equations of motion, to the required orders, are given in Appendices \ref{appA} and \ref{appB}, respectively. If we consider the time component of $T^{\mu \nu}_{\phantom{\mu \nu} ; \nu}=0$ to order $\eta^5$ we see that see that we can write the equation of relativistic energy conservation in the form of the Newtonian equation of mass conservation (\ref{energynm}), as long as we have $p=0$ (i.e. dust). In this case the effective Newtonian variables are as follows:
\be \label{nmmu}
\tilde{\mu} = \mu \left( 1+ \frac{1}{2} v^2 -3 U + \Pi +\nabla^2 E \right) + O(\eta^5)
\ee
and
\be \label{nmv}
\tilde{v}^j = v^j \left( 1 - \frac{1}{2} v^2 + U \right)+ O(\eta^4) \, ,
\ee
where $U$ is the potential defined in Eq. (\ref{appAU}). It is notable that no choice of gauge is yet required in order to put the relativistic energy conservation equation into the form of Eq. (\ref{energynm}), and that the variables $\tilde{\mu}$ and $\tilde{v}^i$ therefore exist in all possible gauges.

The space component of $T^{\mu \nu}_{\phantom{\mu \nu} ; \nu}=0$ to order $\eta^6$ is more complicated, but we find that it can be written in the form in Eq. (\ref{momentum}) if the following is true:
\begin{eqnarray*}
0 &=& -3 v^j \dot{U} - \mathcal{H} v^j v^2 +v^2 U_{,j} - 4 v^j v^k U_{,k} +2 U U_{,j} \\&&
-2 E_{,ij} U_{,i} +2 v^k \dot{E}_{,jk} +v^kv^n E_{,jkn} -2 F_{(i,j)} U_{,i} \\&&+2 v^k \dot{F}_{(j,k)} + v^kv^n F^j_{\phantom{j},nk}
+\phi^{(4)}_{,j} - (\tilde{U}-U)_{,j} \\&&+ \dot{B}_{,j} + \mathcal{H} B_{,j} -\dot{S}_{j} - \mathcal{H} S_j -2 v^k S_{[j,k]} \, ,
\end{eqnarray*}
where we have divided through by a common factor of $\mu$ so that this equation is order $\eta^4$, and where it has been assumed that $h_{ij} = 0=p$. The expression above represents three separate equations, with four degrees of freedom in the choice of gauge. It is expected that all of these equations should be able to be satisfied in many ways (probably infinitely many ways), with one degree of gauge freedom remaining.

Manipulating the above expression, using the solutions to the field equations given in Appendix \ref{appA}, as well as the identities in Appendix \ref{appB}, allows us to write this as the following differential equation:
\be \label{finalnm}
\boxed{
\frac{d^2 {\mathit \Gamma}_j}{d\tau^2} + \mathcal{H} \frac{d {\mathit \Gamma}_j}{d\tau} + U_{,ij} {\mathit \Gamma}_j - \Phi_{7 i,ij} = f }\, ,
\ee
where we have define ${\mathit \Gamma}_j = E_{,j} + F_j$, and where we have introduced the material derivative
\be
\frac{d}{d\tau} = \frac{\partial}{\partial \tau} + v^i \frac{\partial}{\partial x^i}
\ee
and the potential $\Phi_{7i}$, which is defined by
\be
\Phi_{7i} =  - \, a^2(\tau) \int \frac{\mu' {\mathit \Gamma}_i^{\prime}}{\vert {\bf x} - {\bf x'} \vert} \, d^3 x' \, .
\ee
The source function in Eq. (\ref{finalnm}) is a function of the potentials given in Appendix \ref{appA}, such that  $f=f(U,v^i,V^i, \Phi_1, \Phi_2, \delta \Phi_2, \mathscr{A}, \mathscr{B})$, and is given explicitly by
\bea
f &=& - 2 \Phi_{1,j} - 6 \Phi_{2,j} + 5 \delta \Phi_{2,j} + \frac{1}{2} \mathscr{A}_{,j} + \frac{1}{2} \mathscr{B}_{,j} \\ \nonumber
&& -4 \dot{V}_j  - 4 \mathcal{H} V_j - 8 v^i V_{[j,i]} - 3 v^j V_{i,i} \\ \nonumber 
&& -2 \left( U^2 \right)_{,j} - 3 \mathcal{H} v^j U + \mathcal{H}v^j v^2  - v^2 U_{,j} + 4 v^j v^i U_{,i} \, .
\eea
All of the potentials in this expression can be determined from post-processing Newtonian N-body simulations, and in writing $f$ in this way we have chosen to eliminate the vector gravitational potential $W_i$ using the identities in Appendix \ref{appA}.

Putting the metric into Newtonian motion gauge, to first post-Newtonian order, requires choosing a gauge such that Eq. (\ref{finalnm}) is true. Solving this equation will almost certainly have to be done numerically, but once numerical solutions have been obtained then it is clear from Section \ref{pngaugesec} that the gauge can be fixed by a suitable choice of $\xi^i$. This can be seen from Eqs. (\ref{pnEtrans})-(\ref{pnFtrans}). This leaves total gauge freedom in the time component of $\xi^0$, which can be set to any convenient value whilst still maintaining the required property that the equations of motion of test particles obey equations of the same form as they do in Newtonian physics.

Once in this gauge, all relativistic gravitational degrees of freedom can be derived by inverting Eqs. (\ref{nmmu}) and (\ref{nmv}), and then by using the solutions given in Appendix \ref{appA} for the metric perturbations, together with the numerical solutions for $E$ and $F_i$, which can be obtained from ${\mathit \Gamma}_i$. This gives enough information to calculate {\it all} relativistic gravitational effects up to first post-Newtonian order, by post-processing a Newtonian N-body simulation. It is remarkable that this is possible, and that one can in principle obtain a relativistic simulation in this way. We have made no approximations in obtaining this result other than the fluid being dust, which includes the particle interpretation by simply taking the mass density to be $\mu ({\bf x}) = \sum_i m_i \delta({\bf x} - {\bf x}_i)$, for $i$ particles with masses $m_i$ and positions ${\bf x}_i$.

\section{discussion}
\label{secdisc}

We have considered the structure of gauge transformations in both cosmological perturbation theory (applicable on large scales) and post-Newtonian perturbation theory (applicable on small scales). While both treatments of gravitational fields have their own well defined gauge problems, we find that most of the particular gauge choices that are used in cosmology are not valid using post-Newtonian theory in the presence of non-linear structures. In particular, the {\it spatially flat} gauge, the {\it synchronous} gauge, the {\it comoving orthogonal} gauge, the {\it total matter} gauge, the {\it N-body} gauge, and the {\it uniform density} gauge are all beyond the limits of what it is possible to achieve by applying an infinitessimal coordinate transformation in the post-Newtonian sector.

In contrast, the {\it Newtonian motion} gauge appears to be well-defined in both the post-Newtonian and cosmological perturbation theory treatments of gravitational fields, but requires numerical integration of a non-local differential equations (\ref{dvnm2}) and (\ref{finalnm}) in order to be applied in practise. If this is possible, then it should allow one to post-process the results of existing cosmological Newtonian N-body simulations in order to derive relativistic corrections to gravitational fields on all scales, and to determine the effects of these fields on observables {\it without} having to perform additional simulations. This is an intriguing possibility, which we hope to explore further in future studies.

The one standard gauge choice that remains viable, in both of the weak-field treatments that we have considered, is the {\it longitudinal} gauge. The fact that the cosmological perturbation theory equations give sensible results in this gauge, even when the density contrast of matter becomes non-linear, is well known in the cosmology community. Here we formalise this result, and explain its veracity by showing that this gauge is the only commonly used cosmological gauge that can be realised in post-Newtonian expansions (which are purposefully constructed to model weak-field gravity in such situations). This provides support for the use of longitudinal gauge in studies that attempt to simultaneously model both small-scale non-linear structures as well as linear structures on large scales, see e.g. the numerical code {\it gevolution} \cite{gevolution} or the 2-parameter perturbative approach \cite{kit2}.

The fact that one cannot use gauge transformations to change coordinates from a coordinate system that is perturbatively close to FLRW to a synchronous coordinate system in post-Newtonian theory has interesting consequences, but must be interpreted with some care. In particular, this result does {\it not} imply that it is impossible in general to find a coordinate system where the time coordinate corresponds to the proper time of observers comoving with matter (in fact, this is always possible when the matter content is dust \cite{ll}). Instead, it means that the difference between a synchronous coordinate system, and the coordinates of a perturbed FLRW space-time, cannot be related by an infinitesimal gauge generator. That is, the difference between these two different notions of time is large, in the sense defined by the perturbative expansion, and is therefore unattainable by gauge transformations. Such a result would appear to have significance for a number of studies that use proper time in the presence of non-linear structures, such as the calculation of galaxy bias on hypersurfaces of constant proper time \cite{bias}. It may also go someway to explaining the vastly different expectations that different groups of cosmologists appear to have when considering the problem of cosmological back-reaction (see e.g. \cite{clarkson} and \cite{b2}).

\section*{Acknowledgements}

\noindent
We are grateful to Christian Fidler and Marco Bruni for fruitful conversations, and very helpful comments. We acknowledge financial support from the STFC under grant ST/P000592/1.

\appendix

\section{Non-viable choice for Newtonian motion gauge in CPT}
\label{app0}

Instead of Eqs. (\ref{mucpt}) and (\ref{vcpt}), we could have equally well chosen our effective Newtonian variables to be
\be \label{mucpt2}
\tilde{\mu} = \bar{\rho} + \delta \rho - 3 \bar{\rho} \psi  + O(\epsilon^2)
\ee
and
\be \label{vcpt2}
\tilde{v} = v + \dot{E} + O(\epsilon^2) \, ,
\ee
which would have also satisfied the linearized Newtonian equation of energy conservation (\ref{energynmlin}). Substituting into the linearised momentum conservation equation (\ref{momentumnmlin}) from Eq. (\ref{vcpt2}), and taking $\delta \tilde{p}=0$, we find that in this case the following equation must be satisfied:
\be \label{dvnmb}
\dot{v} + \mathcal{H} \, v + \ddot{E} + \mathcal{H} \, \dot{E} = - \tilde{U} \, ,
\ee
where $\hat{U}$ is
\be \label{Unmcptb}
\nabla^2 \hat{U} = 4\pi \, a^2 \,\bar{\rho} \, ( \delta - 3 \psi) \, .
\ee
This can be equivalently written as
\be \label{dvnm2b}
\dot{\sigma} + \mathcal{H} \, \sigma = \phi - \hat{U} \, ,
\ee
where $\sigma = \dot{E}-B$. This equation needs to be satisfied if the Newtonian momentum conservation equation is to be true for the variables in Eqs. (\ref{mucpt2})-(\ref{vcpt2}). 

We can now use the evolution equation for $\sigma$, given below Eq. (\ref{cpt4}), to find that the condition in Eq. (\ref{dvnm2b}) is equivalent to requiring
\be \label{dvnm3b}
\mathcal{H} \, \sigma = \hat{U} - \psi \, .
\ee
In order to evaluate this equation, we can use Eqs. (\ref{cpt1}) and (\ref{Unmcptb}) to write
\be \label{dvnm4b}
\nabla^2 (\tilde{U} - \psi) = \mathcal{H} \, \nabla^2 \sigma - 12 \pi \, a^2 \, \bar{\rho} \left( \psi - \mathcal{H} (v+B) \right) \, .
\ee
This equation makes it clear that Eq. (\ref{dvnm3b}) is satisfied for $\bar{\rho} \neq 0$ if and only if
\be \label{dvnm5b}
\psi - \mathcal{H}(v+B) =0 \, ,
\ee
where sensible boundary conditions have been assumed.

One may now note that the combination of variables on the left-hand side of Eq. (\ref{dvnm5b}) is equal to the curvature perturbation in comoving orthogonal gauge, $\mathcal{R} = \psi - \mathcal{H}(v+B)$. It is therefore impossible to satisfy Eq. (\ref{dvnm5b}), and hence Eq. (\ref{momentumnmlin}), by a choice of gauge using the variables in Eqs. (\ref{mucpt2}) and (\ref{vcpt2}). This shows that the choice of effective variables is extremely important in the implementation of this gauge, and that Newtonian motion gauge {\it cannot} be achieved in every case.

\section{Solving post-Newtonian equations in arbitrary gauge}
\label{appA}

If we write the line-element in as a weak-field perturbation of FLRW, as in Eqs. (\ref{dg00})-(\ref{dgij}), then the leading-order part of the $ij$ field equation can be written as
\bea
a^2 \, R^{(2) i}_{\phantom{(2) i}j} &=& \nabla^2 \psi - (\phi- \psi)_{,ij} + \left( 2 \mathcal{H}^2 + \dot{\mathcal{H}}\right) \delta^i_{\phantom{i} j} - \nabla^2 h_{ij}
\nonumber \\
&=& 4 \pi \, \mu \, a^2 \, \delta^i_{\phantom{i}j}  + \Lambda \, a^2 \, \delta^i_{\phantom{i}j} \, ,
\eea
where the superscript in $R^{(2) i}_{\;\;\;\;\;j}$ indicates that this is the part of this tensor at order $\eta^2$ in the $v/c$ expansion, in appropriately chosen units. This equation immediately tells us that
\be \label{hzero}
\nabla^2 h^{(2)}_{ij} = 0 \qquad \Rightarrow \qquad \boxed{h^{(2)}_{ij} = 0} \, ,
\ee
where appropriate boundary conditions have been used to infer the result on the right. The same equation also gives us
\be \label{f3}
\dot{\mathcal{H}}+2 \mathcal{H}^2 = 4 \pi \, \bar{\mu} \, a^2 + \Lambda \, a^2
\ee
and
\be \label{newU}
\boxed{\phi^{(2)} = \psi^{(2)} = U }\, ,
\ee
where $U$ is the Newtonian gravitation potential that satisfies $\nabla^2 U = 4 \pi \delta \mu \, a^2$, i.e.
\be \label{appAU}
U({\bf x},\tau) = - a^2(\tau) \, \int \frac{\delta \mu({\bf x'},\tau)}{\vert {\bf x} - \bf{x'} \vert} \, d^3 x' \, ,
\ee
where $\delta \mu = \mu - \bar{\mu}$ is the mass density contrast, and appropriate boundary conditions have again been applied.

Next, if we consider the leading-order part of the $0j$ field equation, we find
\bea
-a^2 \, R^{(3) 0}_{\phantom{(3) 0}j} &=& \frac{1}{2} \nabla^2 S_j +\frac{1}{2} \nabla^2 \dot{F}_j + 2 \dot{U}_{,j} +2 \mathcal{H} U_{,j}
\nonumber \\
&=& -8 \pi \mu \, v_j \, a^2 \, ,
\eea
where we have used the results in Eq. (\ref{newU}). Solving this equation we find
\be
\boxed{{S}^{(3)}_j + {\dot{F}_j}^{(2)} = -2 (V_j +W_j)} \, ,
\ee
where the potentials on the right-hand side are given by
\be
V_j = - a^2(\tau) \, \int \frac{\mu({\bf x'},\tau) \, v'_j}{\vert {\bf x} - \bf{x'} \vert} \, d^3 x'
\ee
and
\be
W_j = - a^2(\tau) \, \int \frac{\mu({\bf x'},\tau) \, {\bf v'} \cdot {\bf (x-x')} (x-x')_j}{\vert {\bf x} - {\bf x'} \vert^3} \, d^3 x' \, ,
\ee
and where we have used the result $\dot{U}_{,j} + \mathcal{H} U_{,j} = \frac{1}{2} \nabla^2 (W_j-V_j)$, which can be proven using the continuity equation.

Let us now consider the $00$ field equation. To order $\eta^2$, and using the results above, this equation gives
\be
- 3 \dot{\mathcal{H}} = 4 \pi \, \bar{\mu} \, a^2 - \Lambda \, a^2 \, ,
\ee
which can clearly be seen to correspond to the second Friedmann equation in (\ref{f2}), and which together with Eq. (\ref{f3}) gives the first Friedmann equation (\ref{f1}). The same field equation to order $\eta^4$ gives
\begin{align}
\hspace{-2pt} - a^2 \, R^{(4) 0}_{\phantom{(4) 0} 0} =& \nabla^2 \phi^{(4)} + \nabla^2 {\dot{B}} + \mathcal{H} \nabla^2 B + 3 \ddot{U} 
\nonumber \\&
- U_{,j} \nabla^2 {F}^{}_j  - 2 {F}^{}_{(j,k)} U_{,jk} -2 U_{,j} U_{,j}  
 \nonumber \\&
- \nabla^2 {\ddot{E}} - \mathcal{H} \nabla^2 \dot{E}   - U_{,j} \nabla^2 {E}_{,j}
\nonumber \\&
 - 2 U_{,jk} {E}^{(2)}_{,jk} 
+6 \mathcal{H} \dot{U} + 6 \dot{\mathcal{H}} U 
\nonumber \, .
\end{align}
This result can now be used with the relevant field equation,
\be
- a^2 \, R^{(4) 0}_{\phantom{(4) 0} 0} = 4 \pi \, \mu \, a^2 \left( 2 v^2 + \Pi +3\frac{p}{\mu} \right) \, ,
\ee
to find
\begin{empheq}[box=\fbox]{align*}
&\phi^{(4)}    
+ {\dot{B}}^{(3)} + \mathcal{H} B^{(3)} -U^2 - U_{,j}{F}_j^{(2)} \\[-2pt] &\qquad \qquad \qquad \qquad - {\ddot{E}}^{(2)}- \mathcal{H}{\dot{E}}^{(2)}- U_{,j} {E}_{,j}^{(2)}
\\
\quad =& \frac{1}{2} \Phi_1 + 3 \Phi_2 - 5 \delta \Phi_2 +\Phi_3 +3 \Phi_4\\[-7pt]&\qquad \qquad \qquad \qquad -\delta \Phi_{5j,j} - \delta \Phi_6 +\frac{3}{2} \mathscr{A} + \frac{3}{2} \mathscr{B} \, ,\quad
\end{empheq}
where
\bea
\mathscr{A} &=&  - a^2(\tau) \, \int \frac{\mu'[{\bf v}' \cdot ({\bf x} - {\bf x'})]^2}{\vert {\bf x} - {\bf x'} \vert^3} \, d^3 x'  \nonumber\\
\mathscr{B} &=&  -a^2(\tau) \,  \int \frac{\mu'}{\vert {\bf x} - {\bf x'} \vert} ({\bf x} - {\bf x'}) \cdot \frac{d {\bf v}'}{d\tau}  \, d^3 x'  \nonumber\\
\Phi_1 &=&  - a^2(\tau) \, \int \frac{\mu' \, v^{\prime 2}}{\vert {\bf x} - {\bf x'} \vert}   \, d^3 x'  \nonumber\\
\Phi_2 &=& -a^2(\tau) \,  \int \frac{\mu' \, U'}{\vert {\bf x} - {\bf x'} \vert}   \, d^3 x'  \nonumber\\
\delta \Phi_2 &=& -a^2(\tau)  \,  \int \frac{\delta \mu' \, U'}{\vert {\bf x} - {\bf x'} \vert}   \, d^3 x'  \nonumber\\
\Phi_3 &=& -a^2(\tau) \,  \int \frac{\mu' \, \Pi'}{\vert {\bf x} - {\bf x'} \vert}   \, d^3 x'  \nonumber\\
\Phi_4 &=& -a^2(\tau) \,  \int \frac{p'}{\vert {\bf x} - {\bf x'} \vert}   \, d^3 x'  \nonumber\\
\delta \Phi_{5j} &=& -a^2(\tau) \, \int \frac{\delta \mu' \, {F}_j}{\vert {\bf x} - {\bf x'} \vert}   \, d^3 x'  \nonumber\\
\delta \Phi_{6} &=&  - a^2(\tau) \,\int \frac{\delta \mu'_{,j} {E}_{,j}}{\vert {\bf x} - {\bf x'} \vert}   \, d^3 x' \, . \nonumber
\eea
These are all standard potentials used in post-Newtonian gravity, with the exceptions of $\delta\Phi_2$, $\delta \Phi_{5j}$ and $\delta \Phi_6$, which we have introduced here. Primed quantities in these equations should be taken to mean they are functions of the primed coordinate positions, such that $\mu' = \mu({\bf x}',\tau)$, for example.

In deriving this last result we have used the following identities:
\bea
2 U_{,i} U_{,i} - 8 \pi \bar{\rho} \, U \, a^2 &=& \nabla^2 \left( U^2 - 2 \Phi_2 \right) \nonumber\\
- U_{,j} \nabla^2 {F}_j  - 2 {F}_{(j,k)} U_{,jk} &=& \nabla^2 \left( \delta \Phi_{5j,j} - U_{,j} F_j \right) \nonumber \\
 - U_{,j} \nabla^2 {E}_{,j} - 2 U_{,jk}  {E}_{,jk} &=& -\nabla^2 \left( U_{,j}  {E}_{,j} - \delta \Phi_6 \right) \nonumber \\
\ddot{U} +2 \mathcal{H} \dot{U} + \left( \mathcal{H}^2 +\dot{\mathcal{H}} \right) U &=& -\frac{1}{2} \nabla^2 \left( \mathscr{A} + \mathscr{B} - \Phi_1 \right) \, , \nonumber
\eea
the last of which is proven using the continuity equation. We have also used the following identities in Section \ref{nmsec} of the paper:
\begin{eqnarray*}
\dot{U} + \mathcal{H} U &=& - V_{j,j} = W_{j,j}\\
V_{[j,k]} &=& W_{[j,k]} \\
\dot{V}_{j} - \dot{W}_j &=& - \mathcal{H} \left( V_j - W_j \right) + \mathscr{A}_{,j} + \mathscr{B}_{,j} - \Phi_{1,j} \, .
\end{eqnarray*}
All identities can be proven under the assumption that boundary terms vanish, as would occur (for example) in a space with periodic boundary conditions.

\section{Post-Newtonian equations of motion}
\label{appB}

The equations of motion of post-Newtonian gravity can be obtained by expanding the conservation equations:
\be
T^{\mu \nu}_{\phantom{\mu \nu} ; \nu} =0 \, ,
\ee
which can be conveniently written as
\be \label{con}
\partial_{\nu} \left( \sqrt{-g} \, T^{\mu \nu} \right) + \Gamma^{\mu}_{\phantom{\mu} \rho \nu} \sqrt{-g} \, T^{\rho \nu} =0 \, ,
\ee
where $g$ is the determinant of the metric. The metric in Eqs. (\ref{dg00})-(\ref{dgij}) gives the components of the stress-energy tensor to the required order as
\bea
T^{00} &=& \frac{1}{a^2} \mu \left( 1+ v^2 - 2U +\Pi \right) + O(\eta^5) \nonumber\\
T^{0i} &=& \frac{1}{a^2} \mu \, v^i \left( 1+ \frac{1}{2} v^2 - U +\Pi \right) + \frac{1}{a^2} p \, v^i +O(\eta^6) \nonumber\\
T^{ij} &=& \frac{1}{a^2} \left( \mu v^i v^j + p \delta^{ij} \right) \hspace{3cm} +O(\eta^7)
\nonumber\\[-4pt]&&+ \frac{1}{a^2} \left[ (\mu \Pi + p)v^i v^j + 2U p \delta^{ij} - 2 (E_{,}^{\phantom{,}i} +F^i ) p \right] \nonumber \, .
\eea 
Likewise, the connection coefficients, up to the required order, are given by
\begin{eqnarray*}
\Gamma^0_{\phantom{0} 00} &=& H + \dot{U} + O(\eta^4) \\
\Gamma^0_{\phantom{0} 0i} &=&  U_{,i} + O(\eta^3)   \\
\Gamma^0_{\phantom{0} ij} &=& \delta_{ij} \mathcal{H} + O(\eta^2)    \\
\Gamma^j_{\phantom{j} 00} &=&  U_{,j} + \phi^{(4)}_{,j} + \dot{B}_{,j} + \mathcal{H} B_{,j} -\dot{S}_j - \mathcal{H} S_j \\&&+2 U U_{,j} - 2 E_{,ij} U_{,i} - 2 F_{(i,j)} U_{,i} +O(\eta^5) \\
\Gamma^j_{\phantom{j} 0k} &=&  \delta_{jk} ( \mathcal{H} - \dot{U}) - S_{[j,k]} + \dot{E}_{,jk} +\dot{F}_{(j,k)} +O(\eta^4)  \\
\Gamma^j_{\phantom{j} kn} &=&   -\delta_{jn} U_{,k} -\delta_{jk} U_{,n} +\delta_{kn} U_{,j} \\&&+E_{,jkn} +F^j_{\phantom{j} ,nk} + O(\eta^3) \, ,
\end{eqnarray*}
and the square root of the determinant of the metric is
\be
\nonumber
\sqrt{-g} = a^4 (1-2 U+ \nabla^2 E ) + O(\eta^3)  \, .
\ee
In deriving all of these equations we have used the results from Eqs. (\ref{hzero}) and (\ref{newU}) to eliminate $h^{(2)}_{ij}$, and to write $\phi^{(2)}$ and $\psi^{(2)}$ in terms of $U$.

The order $\eta^3$ part of the time component of Eq. (\ref{con}) can immediately be seen to reproduce the Newtonian equation of mass conservation on an expanding background, as given in Eq. (\ref{energy}). Likewise, the order $\eta^4$ of the spatial components of Eq. (\ref{con}) gives the momentum conservation equation from (\ref{momentum}), once we set $\phi=U$. The next non-vanishing contributions to the conservation equations (\ref{con}) come at order $\eta^5$ in the time component, and order $\eta^6$ in the space components. These correspond to first post-Newtonian order, in the normal language of this type of weak-field expansion (though on a cosmological background here). We will now consider each of these in turn.

\begin{widetext}
Calculating the order $\eta^5$ part of the time component of Eq. (\ref{con}), and simplifying using the momentum conservation equation (\ref{momentum}), gives
\bea
0 &=& \partial_{\tau} \Bigg[ a^3 \mu \left( \frac{1}{2} v^2 -3 U + \Pi + \nabla^2 E \right) \Bigg] 
+ \partial_j \Bigg[ a^3 \mu \, v^j \left( -2 U + \Pi + \nabla^2E \right) \Bigg] 
+a^3 p \left (v^j_{\phantom{j} ,j} + 3 \mathcal{H} \right) \, .
\eea
Calculating the order $\eta^6$ part of the spatial components of Eq. (\ref{con}), and taking $p=0$, gives
\begin{eqnarray*}
0 &=& \partial_{\tau} \Bigg[ a^4 \mu v^j \left( \frac{1}{2} v^2 -3 U+\Pi +\nabla^2 E\right) \Bigg] +\partial_{k} \Bigg[a^4 \mu v^j v^k \left( -2 U+\Pi +\nabla^2 E\right) \Bigg]\\
&& + a^4 \mu \Bigg[ U_{,j} \left( 2 v^2-4 U+\Pi +\nabla^2 E \right) + \phi^{(4)}_{,j} + \mathcal{H} B_{,j} + \dot{B}_{,j} - \mathcal{H} S_{j} - \dot{S}_{j} +2 U U_{,j} -2 E_{,ij} U_{,i} -2 F_{(i,j)} U_{,i} \Bigg]\\
&&-2 a^4 \mu v^k \Bigg[ S_{[j,k]} + \delta_{jk} \dot{U} - \dot{E}_{,jk} -\dot{F}_{(j,k)} + v^j U_{,k} - \frac{1}{2} v^n E_{,jkn} -\frac{1}{2} v^n F^j_{\phantom{j},kn} \Bigg] \, .
\end{eqnarray*}
These are all of the equations that are required to calculate the trajectories of test particles to first post-Newtonian order.

\vspace{1cm}
{\phantom{1}}
\end{widetext}

\end{document}